\documentclass[doublecol]{epl2}
\usepackage{amsmath}
\usepackage{bm}
\newcommand{\nwl}{\\begin{equation}2mm]}
\newcommand{\edc}{\end{document}}
\newcommand{\bb} {\begin{thebibliography}}
\newcommand{\eb}{\end{thebibliography}}

\newcommand{\bc}{\begin{center}}
\newcommand{\ec}{\end{center}}

\newcommand{\be}{\begin{equation}\small}
\newcommand{\ee}{\end{equation}\normalsize}
\newcommand{\bea}{\begin{eqnarray}}
\newcommand{\eea}{\end{eqnarray}}
\newcommand{\ba}{\begin{array}{l}   }
\newcommand{\lab}[1]{\label{#1}}
\newcommand{\ea}{\end{array}}

\newcommand{\re}[1]{(\ref{#1})}

\newcommand{\ci}{\cite}

\newcommand{{\vergul}}{  ,}

\newcommand{\wh}[1]{{\widehat{#1}}}

\title{Interference of spin-orbit coupled Bose-Einstein condensates.}
\shorttitle{Interference of SOC BEC} 

\author{Sh. Mardonov\inst{1,5,6}\and M. Palmero\inst{1} \and M. Modugno\inst{2,4} \and E. Ya. Sherman\inst{1,2}  \and J. G. Muga\inst{1,3}}
\shortauthor{Sh. Mardonov\etal}

\institute{
  \inst{1} Department of Physical Chemistry, The University of the Basque Country UPV/EHU, 48080 Bilbao, Spain\\
  \inst{2} IKERBASQUE Basque Foundation for Science, Bilbao, 48011 Bizkaia, Spain\\
  \inst{3} Department of Physics, Shanghai University, 200444 Shanghai, People's Republic of China\\
  \inst{4} Dpto. de F\'{i}sica Te\'{o}rica e Hist. de la Ciencia, Universidad del Pa\'{i}s Vasco UPV/EHU, 48080 Bilbao, Spain\\
  \inst{5} The Samarkand Agriculture Institute, 140103 Samarkand, Uzbekistan\\
  \inst{6} The Samarkand State University, 140104 Samarkand, Uzbekistan

  }

\abstract{
Interference of atomic Bose-Einstein condensates, observed in free expansion experiments,
is a basic characteristic of their quantum nature. The ability to produce
synthetic spin-orbit coupling in Bose-Einstein condensates has recently opened a new research field.
 {Here we theoretically describe interference of two noninteracting
spin-orbit coupled Bose-Einstein condensates in an external synthetic magnetic field.
We demonstrate that the spin-orbit and the Zeeman couplings strongly
influence the interference pattern determined by the angle between the spins of the
condensates, as can be seen in time-of-flight experiments.} We show that a
quantum backflow, being a subtle feature of
the interference, is, nevertheless, robust against the spin-orbit coupling and applied synthetic magnetic field.}

\begin{document}

\maketitle

\section{I. Introduction} Interference of matter waves is one
of the most interesting effects in quantum physics.  The interference of two
expanding Bose-Einstein condensates is a clear manifestation
of quantumness in macroscopic systems \ci{castin1997, andrews, band}.
It can be observed by preparing two condensates in spatially separated
harmonic traps, that are released afterwards. Then, the condensates
can expand freely and eventually overlap, producing an interference pattern.

The quantum dynamics becomes much richer for spin-orbit coupled Bose-Einstein
condensates,  where optically produced pseudospin is coupled to the atomic momentum 
and to a synthetic, also optically produced, magnetic field \ci{spielman2009,wang2010,spielman2011}.  
These effects, which open a venue to the simultaneous control of orbital
and  spin degrees of freedom {and to experimental observation of new phases and dynamic 
processes} have been discussed for a variety of ultracold atomic
systems \ci{artem, osterloh, ruseckas, galitski, liu, dalibard, ho2011, stringari,anderson, zhang2012, zhang}
including recently produced and studied Fermi  gases with synthetic spin-orbit coupling \cite{wang2012,cheuk2012}.
 {In quantum information technologies, spin-orbit coupled Bose-Einstein condensates can serve 
as a realization of macroscopic qubits, as proposed in \cite{galitski}.}
State-of-the-art reviews can be found in \cite{zhai2012,spielman2013}.

\begin{figure}[t]
\begin{center}
\includegraphics[height=5cm,width=7cm]{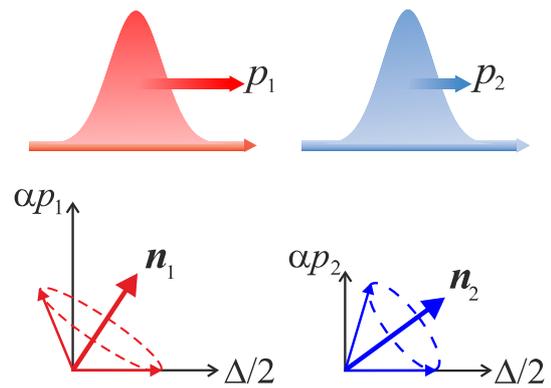}
\caption{Two condensates with mean momenta per particle $p_{1}$, $p_{2}$ and spin-orbit
coupling constant $\alpha$ and spins precessing in a synthetic magnetic field characterized
by Zeeman splitting $\Delta$. Dashed ellipses show the time-dependent spins of condensates,
and vectors ${\bm n}_{1}$ and ${\bm n}_{2}$ defined in Eq.\re{3.28} mark corresponding precession axes.}
\label{packets}
\end{center}
\end{figure}
 {Here we consider time-of-flight control 
of interference of two spin-orbit- and Zeeman-coupled one-dimensional condensates (as shown in Fig.\ref{packets}) 
producing their entangled state, 
which might be required for quantum information purposes \cite{galitski}.}
The condensates, that move freely in a waveguide
realized by tight confinement in the transverse directions,
give rise to an interference pattern that strongly
depends on the relative orientation of their pseudospins.
We study the role of the synthetic magnetic field on the interference and show
that it can be fully controlled by changing the synthetic Zeeman coupling.
In addition, we show that the quantum backflow
\ci{al, bracken, gonsalo-sala, gonsalo-leavens, damborenea, berry, yersley,mikel},
being a subtle effect of the interference, is rather
robust against mutual orientation of spins of the condensates.

\section{ {II. Interference of condensates with spin-orbit and Zeeman coupling}}

 {To study the spin-dependent interference of two condensates, 
we take the synthetic magnetic field along the $x$-axis 
and spin-orbit coupling field along the $z$- axis. The Hamiltonian becomes:
\bea
\widehat{H}=\frac{\widehat{p}^{2}}{2M}+\frac{\alpha}{\hbar}\widehat{p}\widehat{\sigma}_{z}
+\frac{\Delta}{2}\widehat{\sigma}_{x},
\lab{3.27}
\eea
where $\widehat{p}$ is the momentum operator, $M$ is the particle mass, $\alpha$ is the spin-orbit coupling constant,
$\widehat{\sigma}_{z}$ and $\wh{\sigma}_{x}$ are the Pauli matrices, and $\Delta$ is the Zeeman splitting.
To see the qualitative effect of the spin-orbit coupling, we begin with
a single packet where the solution of the Schr\"{o}dinger equation
\begin{equation}
i\hbar\frac{\partial{\bm \Psi}}{\partial t}=\widehat{H}{\bm \Psi},\quad 
{\bm \Psi}\equiv{\bm \Psi}(x,t)\equiv\left
[\begin{array}{c}\psi^{\uparrow}(x,t)\\ \psi^{\downarrow}(x,t)\end{array}\right]
\lab{1.2}
\end{equation}
is 
\begin{equation}
{\bm \Psi}(x,t)=\int \exp\left(-\frac{i}{\hbar}\widehat{H}t+\frac{i}{\hbar}px\right)\mathbf{G}(p)\frac{dp}{2\pi\hbar}.
\lab{1.3}
\end{equation}
Here $\mathbf{G}(p)=g(p){\bm B}(0)$, $g\left(p\right)$ is the 
wave function in the momentum space, and
${\bm B}(0)=\left(\beta_{1}(0),\ \beta_{2}(0)\right)^{T}$ is the
initial spinor normalized with $\beta_{1}^{2}(0)+\beta_{2}^{2}(0)=1$.
We assume without loss of generality that $\beta_{1}(0)$ and $\beta_{2}(0)$
are real.  For definiteness, we take Gaussian $g(p)$, produced by an 
initial state in a harmonic trap, as described in Ref.\cite{mikel}. This function is given by:
\begin{equation}
g(p)=(4\pi w^2)^{1/4}\exp\left[-\frac{{w}^2(p-\langle p\rangle)^2}{2\hbar^2}-\frac{ipx_{\rm in}}{\hbar}\right],
\lab{2.11a}
\end{equation}
where $\langle p\rangle$ is the mean momentum, $w$ is the initial width, and $x_{\rm in}$
is the initial position.

For a packet narrow in the momentum space with $\langle p\rangle w\gg\hbar,$
one can neglect momentum distribution and write the wavefunction \re{1.3}
as a product ${\bm\Psi}(x,t)=\psi(x,t){\bm B}(t),$ where $\psi(x,t)$ is the time and coordinate
dependence in the absence of spin-orbit coupling. The spin state of a packet is given by:
\bea
{\bm B}(t)=\exp\left(-\frac{i}{\hbar}\widehat{H}_{s}t\right){\bm B}(0),
\lab{2.17}
\eea
where ${H}_{s}=\alpha \langle p\rangle\widehat{\sigma}_{z}/\hbar + \Delta\sigma_{x}/2$
is the mean value of a spin contribution to the Hamiltonian \re{3.27}. 
We use Eq.(\ref{2.17}) below for a qualitative analysis of 
the packets' interference.

As a result, the spin of a wavepacket with well-defined 
momentum $\langle p\rangle$ rotates around the axis (cf. Fig.\ref{packets}):
\bea
\bm{n}=\frac{2}{\hbar\Omega}\left[\frac{\Delta}{2},\ 0,\ \frac{\alpha \langle p\rangle}{\hbar} \right],
\lab{3.29}
\eea
with the rate
\bea
\Omega=\frac{2}{\hbar}\sqrt{\left(\frac{\alpha\langle p\rangle}{\hbar}\right)^2+\left(\frac{\Delta}{2}\right)^2},
\lab{3.28}
\eea
where $\hbar\Omega\left(\bm{n}\cdot{\bm\sigma}\right)=2H_{s}.$
Figure \ref{packets} shows rotating spins 
in the presence of Zeeman splitting and directions of the vectors ${\bm n}_{j}$ in \re{3.29}, 
where index $j=1,2$ labels the condensate, and we use $p_{j}$ for corresponding mean
values. 

If $\Delta=0$, the spinor components in \re{1.3} are decoupled and have the form:
\bea
&&\hspace{-1cm}\psi^{\uparrow}(x,t)=\beta_{1}(0)\int g(p)
\exp\left[-\frac{i}{\hbar}\frac{p^{2}t}{2M}+\frac{ipx_{-}}{\hbar}\right]\frac{dp}{2\pi\hbar},
\nonumber
\\
&&\hspace{-1cm}\psi^{\downarrow}(x,t)=\beta_{2}(0)\int g(p)
\exp\left[-\frac{i}{\hbar}\frac{p^{2}t}{2M}+\frac{ipx_{+}}{\hbar}\right]\frac{dp}{2\pi\hbar},
\lab{1.5}
\eea
where $x_{\pm}\equiv x\pm\alpha t/\hbar.$ The $\alpha$-determined phase shift between
$\psi^{\uparrow}(x,t)$ and $\psi^{\downarrow}(x,t)$ in \re{1.5} leads to a coordinate-dependent 
spin rotation. The same results for spin motion can be obtained by
gauging out the spin-orbit coupling in Eq.\re{3.27} by a coordinate-dependent spin rotation $\exp[i\sigma_{z}x/L_{\rm so}]$ 
($L_{\rm so}\equiv \hbar^{2}/M\alpha$ is the spin rotation length),
calculating  the resulting dynamics, and then making  the inverse transformation to obtain the 
observables \cite{Tokatly,Glazov}. 
However, in the presence of a Zeeman field, which is of our interest, gauging out the spin-orbit coupling leads 
to a coordinate-dependent effective magnetic field. Although transport effects
can be obtained with Eq.\re{2.17} (see, e.g \cite{Pershin}), general dynamics 
is difficult to treat beyond perturbation theory \cite{Levitov}. 
For this reason we use the direct calculation rather than the spin rotation approach.}

The expectation values of the packet width and velocity at time $t$ obtained with Eq.\re{1.5} are
\bea
\left\langle w(t)\right\rangle =\left[w^{2}\left(1+\frac{\hbar^{2}t^{2}}{w^{4}M^{2}}\right)
+8\frac{\alpha^{2} t^{2}}{\hbar^{2}}
\beta_{1}^{2}(0)\beta_{2}^{2}(0)\right]^{1/2},
\lab{1.7}
\eea
\bea
\left\langle v\right\rangle \equiv \frac{i}{\hbar}\langle[\wh{H},\wh{x}]\rangle =
\frac{\langle p\rangle}{M}+\frac{\alpha}{\hbar}\left\langle\widehat{\sigma}_{z}(0)\right\rangle.
\lab{1.8}
\eea
Here $\left\langle \widehat{\sigma}_{z}(0)\right\rangle ={\bm B}^{\dag}(0) \widehat{\sigma}_{z} {\bm B}(0) =
\beta_{1}^{2}(0)-\beta_{2}^{2}(0)$.

For a general form of ${\bm \Psi}(x,t)$ the current density $J(x,t)$ is given by:
\be
J(x,t)=\frac{i\hbar}{2M}\left[ {\bm \Psi}_{x}^{\dagger}{\bm \Psi}-{\bm \Psi}^{\dagger}{\bm \Psi}_{x}\right]
+\frac{\alpha}{\hbar}{\bm \Psi}^{\dagger}\widehat{\sigma}_{z}{\bm \Psi},
\lab{1.10}
\ee
where ${\bm \Psi}_{x}\equiv{\partial}{\bm \Psi}/{\partial x}.$
The experimentally measured density $\rho(x,t)={\bm \Psi}^{\dagger}(x,t){\bm \Psi}(x,t)$
is related to $J(x,t)$ by the continuity equation. For the weak coupling considered below
we neglect the $\alpha$-related terms in Eqs.\re{1.7}-\re{1.10}.

To see the effect of the spin-orbit coupling on interference
of condensates, we take the initial wave function in the form {where the coherence can be achieved,
e.g. by a technique proposed in \cite{castin1997}}:
\be
\mathbf{G}(p)=A_{1}g_{1}(p){\bm B}_{1}(0)+A_{2}g_{2}(p){\bm B}_{2}(0).
\lab{2.11}
\ee
Here the amplitudes $A_{1}$ and $A_{2}$ are normalized as
$A_{1}^{2}+A_{2}^{2}=1,$ ${\bm B}_{j}(0) =\left(\beta_{j1}(0),\beta_{j2}(0)\right)^{T}$
is the corresponding spinor, and $g_{j}(p)$ is defined by \re{2.11a}.
The average velocities of the packets $v_{j}$ are determined by \re{1.8}
for the corresponding momentum $p_{j}$ and spin state, and from now on we omit $\langle\ldots\rangle$
in the notation of averages. Using \re{1.3} and \re{2.11}
we obtain the exact evolution of two initial wave packets with spin-orbit coupling.

For a qualitative understanding we use a model of two independent condensates moving
with different momenta. Take first as an illustration a system with the following 
${\bm \Psi}(x,t)$:
\bea
{\bm \Psi}(x,t) ={\bm \Psi}_{1}(x,t)+{\bm \Psi}_{2}(x,t),
\lab{2.12}
\eea
where
\bea
{\bm \Psi}_{j}(x,t) =\psi_{j}(x,t){\bm B}_{j}(t).
\lab{2.13}
\eea
The current density \re{1.10} for the wave function \re{2.12} is defined by:
\bea
J(x,t)=
&&\hspace{-.5cm}\frac{\hbar}{M}\Im \left[ \psi_{1}^{\dagger}\frac{\partial\psi_{1}}{\partial x} |{\bm B}_{1}|^{2}
+ \psi_{1}^{\dagger}\frac{\partial\psi_{2}}{\partial x} {\bm B}_{1}^{\dagger}{\bm B}_{2}\right]+
\nonumber
\\
&&\hspace{-.5cm}\frac{\hbar}{M}\Im \left[\psi_{2}^{\dagger}\frac{\partial\psi_{1}}{\partial x} {\bm B}_{2}^{\dagger}{\bm B}_{1}
+ \psi_{2}^{\dagger}\frac{\partial\psi_{2}}{\partial x} |{\bm B}_{2}|^{2} \right],
\lab{2.14}
\eea
where ${\bm B}_{1}\equiv{\bm B}_{1}(t)$ and ${\bm B}_{2}\equiv{\bm B}_{2}(t)$. Equation \re{2.14}
shows that the interference, seen here as the fast oscillations
in the coordinate or time-dependence of the current, is controlled by the spin states through the 
product ${\bm B}_{1}^{\dagger}{\bm B}_{2}.$

 {General expressions for ${\bm B}_{1}^{\dagger}(t){\bm B}_{2}(t)$ are cumbersome. 
Taking as an example packets with well-defined momenta and spins initially parallel to the 
$x$-axis, we find with \re{2.17}:
\begin{eqnarray}
&&{\bm B}_{1}^{\dagger}(t){\bm B}_{2}(t) = \nonumber \\
&&{\frac{\hbar^{2}\Delta^2+4p_{1}p_{2}\alpha^2}{\hbar^4\Omega_{1}\Omega_{2}}} \sin\frac{\Omega_{2}t}{2} \sin\frac{\Omega_{1}t}{2}+
\cos\frac{\Omega_{2}t}{2} \cos\frac{\Omega_{1}t}{2}+\nonumber\\
&&i\left[\frac{\Delta}{\hbar\Omega_{2}} \sin\frac{\Omega_{2}t}{2} \cos\frac{\Omega_{1}t}{2}-
\frac{\Delta}{\hbar\Omega_{1}} \cos\frac{\Omega_{2}t}{2} \sin\frac{\Omega_{1}t}{2}\right],
\lab{3.31}
\end{eqnarray}
where $\Omega_{1}$ and $\Omega_{2}$ are defined by \re{3.28}.
If $\Delta=0,$ we obtain ${\bm B}_{1}^{\dagger}(t){\bm B}_{2}(t)=\cos\theta(t)$,
with the angle 
\bea
\theta(t)=\frac{\alpha}{\hbar^{2}}\left(p_{1}-p_{2}\right)t.
\label{2.19}
\eea
Equation \re{3.31} shows how the mutual orientation of the spins of the condensates 
and, in turn, their interference, depends on the time of flight in the presence of spin-orbit or Zeeman coupling.
In particular, if at $t=t_{0}$ the spin states are orthogonal,
the interference disappears, which shows that it can be controlled by manipulating the condensate spin.}

\begin{figure}[t]
\begin{center}
\includegraphics[height=4cm,width=7cm]{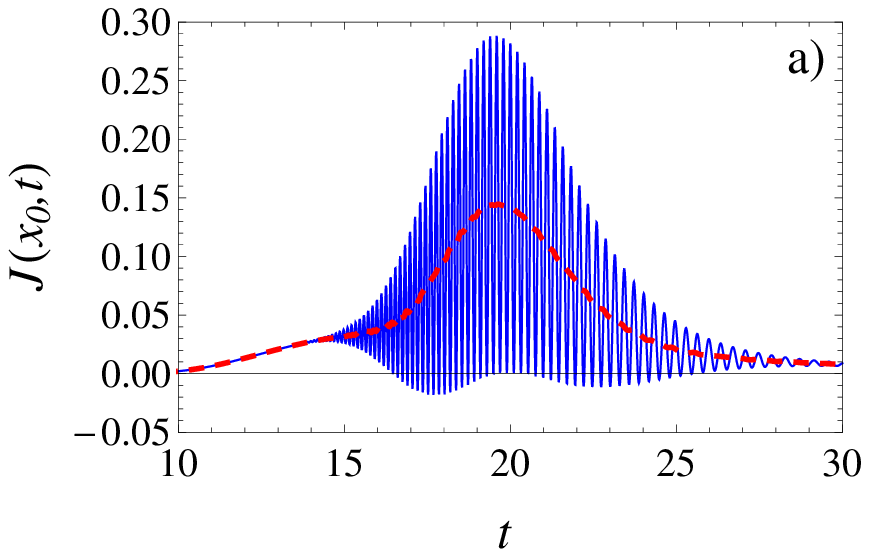}
\includegraphics[height=4cm,width=7cm]{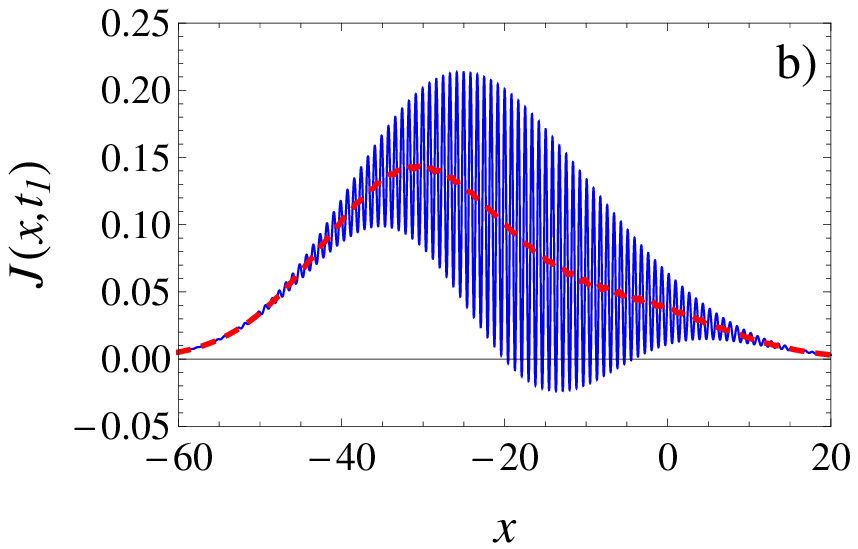}
\caption{Plot of current  (a) vs time at $x=x_{0}$ and (b) vs coordinate at $t=t_{1}\ (t_{1}=16)$,
for the parameters in \re{2.20}.
Color lines correspond to values of spin-orbit coupling
from \re{2.21a} and \re{2.21}, $\alpha_{1}$ - red dashed line, $\alpha_{2}$ - blue solid line.}
\label{destroy}
\end{center}
\end{figure}

Let us now consider specific examples for $\Delta=0$. We use in numerical calculations
the system of units with $\hbar \equiv 1$, mass $M \equiv 1$, unit length of one micron,  
and dimensionless parameters:
\begin{eqnarray}
&&p_{1}=8,\ p_{2}=2,\ A_{1}=A_{2}=1/\sqrt{2},\nonumber\\
&&\beta_{11}(0)=\beta_{12}(0)=\beta_{21}(0)=\beta_{22}(0)=1/\sqrt{2},
\lab{2.20}
\end{eqnarray}
corresponding to both condensates with the spin oriented along the $x$-axis.
We take time-of-flight $t_{0}=20$ and the wave packet ``collision'' point at $x_{0}=0$.
To make connection with possible experimental observations, we take 
$^{87}\mbox{Rb}$ atom as an example. The resulting
velocity unit $\hbar/(M_{\rm Rb}\times 10^{-4}\mbox{ cm})$ is 0.072 cm/s
and, therefore, the unit of time is approximately $1.4\times10^{-3}$ s.
As a result, $t_{0}=20$ corresponds to about
28 milliseconds and the initial distance between the
packets (for $p_1=8$ and $p_2=2$) of 120 microns. Below we consider two
realizations of the condensates, with equal and different widths.

$\mathbf{1.}$ First, we take both initial widths equal, $w_{1}=w_{2}=1$.
At the meeting time $t_{0}$, if the spin states of the condensates are orthogonal,
that is $\cos\theta(t_{0})=0,$ the interference is destroyed.
For $\cos\theta(t_{0})=\pm1$, we obtain the constructive (destructive)
interference with similar fringes, just shifted
by half period. Here the interference
is maximal and, for $\cos\theta(t_{0})=1,$ the same as
in the absence of spin-related effects.
We take spin-orbit coupling corresponding to the two realizations of the angle 
between the spins of condensates at the collision point (see \re{2.19}):
\bea
&&\alpha_{1}=\frac{\pi\hbar^{2}}{2(p_{1}-p_{2})t_{0}},\qquad \theta(t_{0})=\pi/2
\lab{2.21a}
\\
&&\alpha_{2}=\frac{\pi\hbar^{2}}{(p_{1}-p_{2})t_{0}}, \qquad \theta(t_{0})=\pi
\lab{2.21}
\eea
respectively.
\begin{figure}[t]
\begin{center}
\includegraphics[height=4cm,width=7cm]{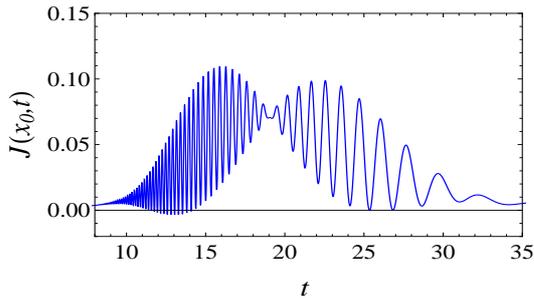}
\caption{Plot of current density for $p_{1}=5$, $p_{2}=2.5,$ $\alpha=0.09,\ w_{1}=0.5,
\ w_{2}=20$ and other parameters from \re{2.20}.
For these parameters the number of no-interference points in \re{2.22} is $N_{\cos\theta=0}\approx 1$.}
\label{periodic}
\end{center}
\end{figure}

In Fig. \ref{destroy} one can see that, for $\alpha=\alpha_{1}$ there is no interference in the flux, while
for $\alpha=\alpha_{2}$ the flux is characterized by a strong interference pattern
and by the presence of backflow, namely a negative current density,
$J(x_{0},t)<0$, see Fig. \ref{destroy} \ci{al, bracken, gonsalo-sala, gonsalo-leavens, damborenea, berry, yersley,mikel}.
For other values of spin-orbit coupling the flux interference is between these two
limits. 

\textbf{2.} 
For different initial widths of the packets 
at the time-of-flight $t_{0}$ the spreads of the packets
can be, in general, different. The same holds for the travel time
of the packets through the point $x_{0}$ defined as
\bea
T_{j}\approx 2\frac{w_{j}(t_{0})}{ v_{j}},
\lab{2.24a}
\eea
where $ v_{1}$ and $ v_{2}$ are the velocities of the packets
determined in \re{1.8},  $ w_{1}(t_{0})$ and $ w_{2}(t_{0})$ are
the widths of the packets at the meeting time $t_{0}$ determined in \re{1.7}, and we have taken into account
that the traveling time of wave function is of the order of $2w_{j}(t)$.

In this case, the duration of the interference (interference time) is
\bea
T_{\rm int}\approx \min\{T_{1},\ T_{2}\}.
\lab{2.24}
\eea

From \re{2.19} we define one rotation period $T_{\rm rot}$ as 
\bea
\theta(T_{\rm rot})=2\pi
\lab{2.23a}
\eea
and obtain
\bea
T_{\rm rot}=\frac{2\pi\hbar^{2}}{\alpha(p_{1}-p_{2})}.
\lab{2.23}
\eea

If the coupling $\alpha$ is large, the spins of the
packets rotate fast and 
during the time interval $T_{\rm int}$ the interference would be
destroyed several times depending on the rotation rate.
Then the number of points in time domain where the interference disappears can be estimated as:
\bea
N_{\cos\theta=0}\approx2\frac{T_{\rm int}}{T_{\rm rot}},
\lab{2.22}
\eea
In this formula the factor 2 means that, in one period of rotation of the angle
between spins, the interference is destroyed twice when spin states are orthogonal.
In addition, if the packets are initially narrow, and, therefore, spread with a large
rate of the order of $\hbar/Mw_{j}$, one can see the effect of multiple interferences better.
From Fig. \ref{periodic} one can see that, during the interference time
the spin states become orthogonal once, and the interference is destroyed
at this instant.
\begin{figure}[t]
\bc
\includegraphics[height=4cm,width=7cm]{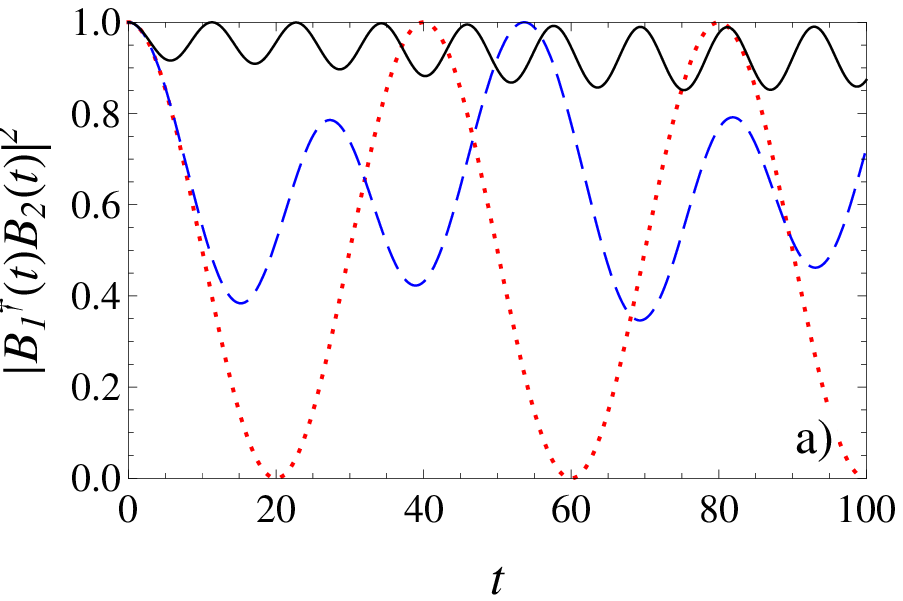}
\includegraphics[height=4cm,width=7cm]{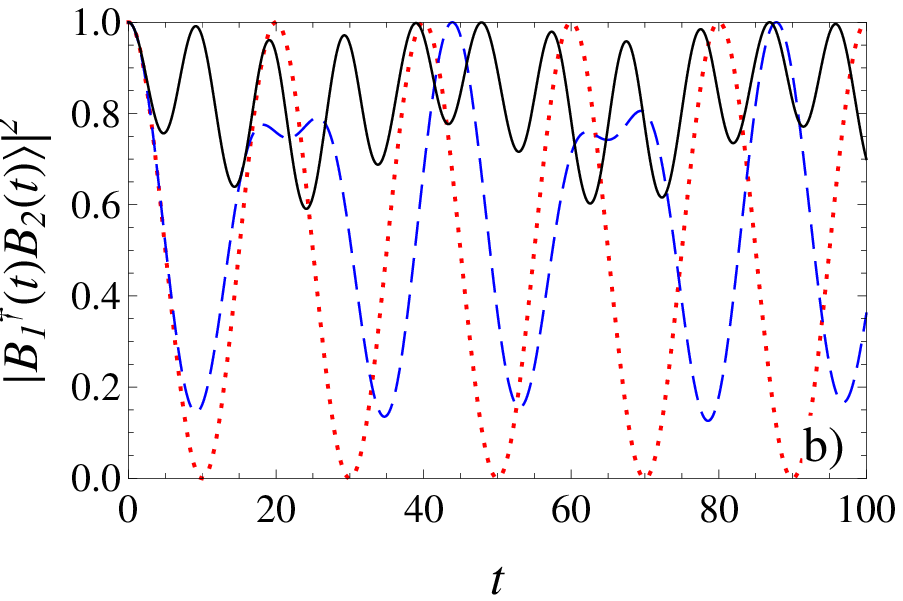}
\caption{Products of spinors for values in \re{2.20}
for the spin-orbit coupling from \re{2.21a} and \re{2.21}: (a) - $\alpha_{1}$ and (b) - $\alpha_{2}$.
Lines correspond to: $\Delta=0$ - red dot line, $\Delta=0.1$ -
blue dashed line, and $\Delta=0.5$ - black solid line.}
\label{precessions}
\ec
\end{figure}

\begin{figure}[t]
\bc
\includegraphics[height=4cm,width=7cm]{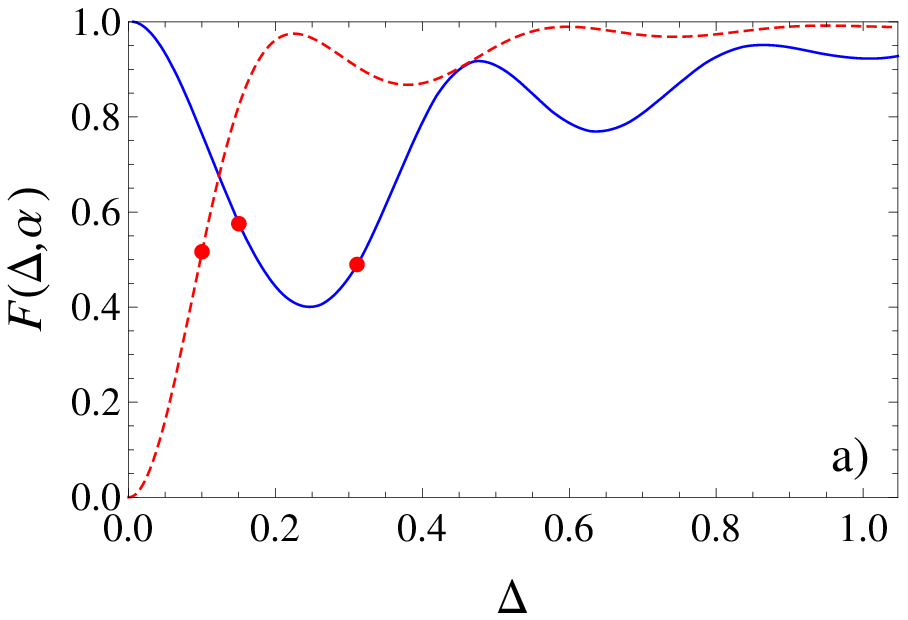}
\includegraphics[height=4cm,width=7cm]{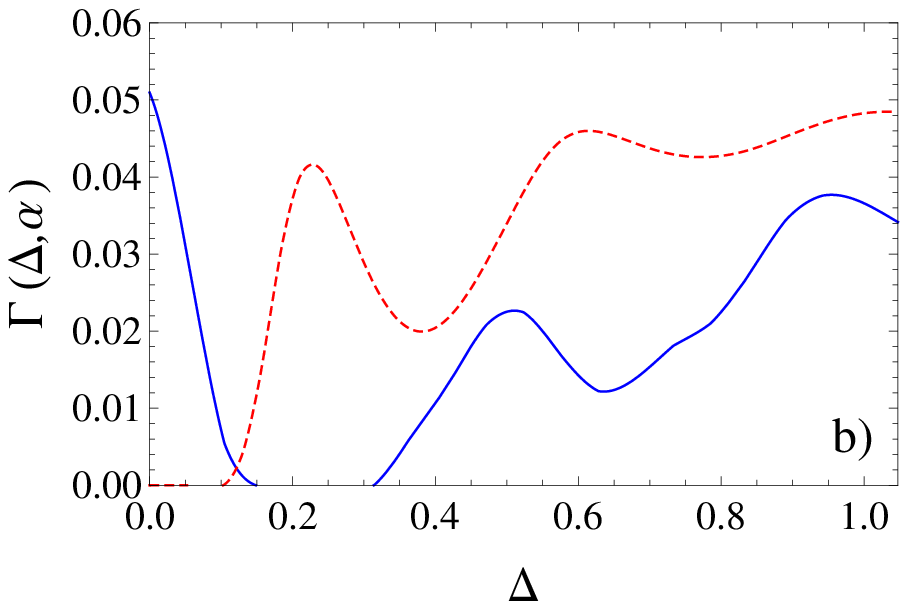}
\caption{The evaluation of interferences (a) and backflow (b) for values from \re{2.20}, $w_{1}=w_{2}=1$
and spin-orbit coupling from \re{2.21a} and \re{2.21} $\alpha_{1}$-dashed red line,
$\alpha_{2}$-solid blue line. In (a) small circles mark appearance and disappearance of the backflow.}
\label{interference}
\ec
\end{figure}

Below we consider the effects of the Zeeman term limiting ourselves to 
equal initial widths of the packets with all other initial parameters \re{2.20}
unchanged. It is important here that if the condition $\Delta\gg\alpha p_{j}/\hbar$ is satisfied, 
the vectors \re{3.29} are very close to each other and to the $x$-axis, and the spins
be always parallel to each other with a high accuracy.
Fig. \ref{precessions} demonstrates the spin states \re{3.31} and shows that
the expression \re{3.31} is zero only when $\Delta=0$. As a result with synthetic magnetic field the interference
cannot be completely destroyed by spin-orbit coupling.
\begin{figure}[t]
\bc
\includegraphics[height=4cm,width=7cm]{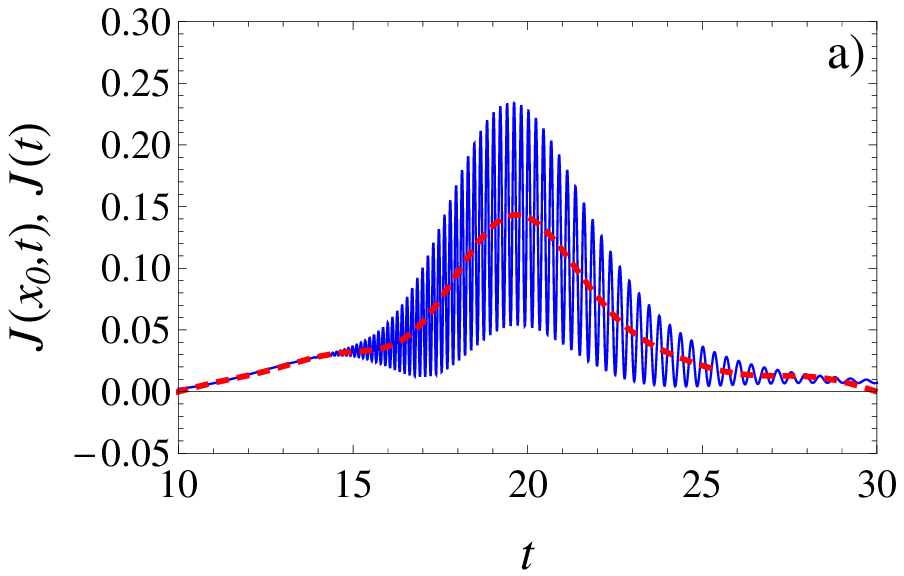}
\includegraphics[height=4cm,width=7cm]{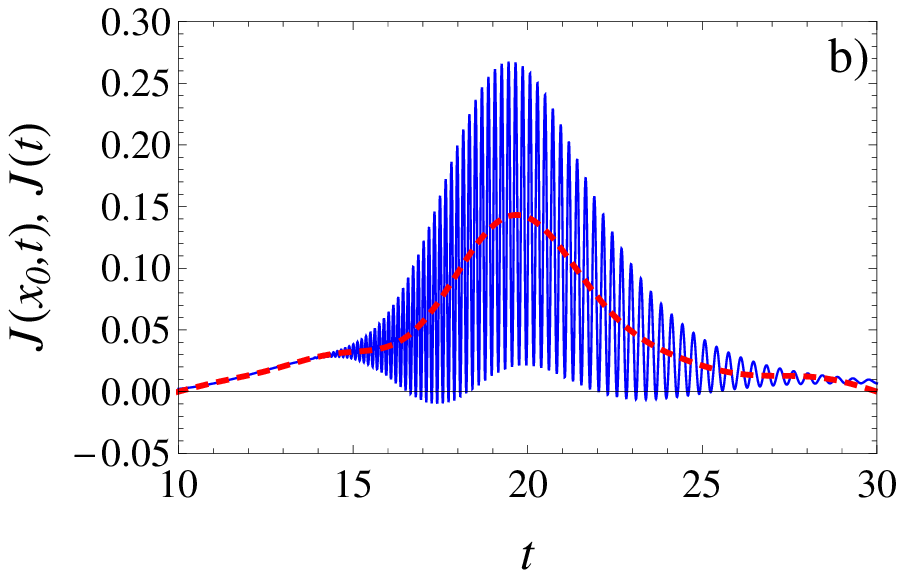}
\caption{The fluxes for values from \re{2.20}, spin-orbit coupling is $\alpha_{2}$ from \re{2.21},
and the plots correspond to the fields with $\Delta=0.26$ (a) and $\Delta=0.63$ (b).
Lines correspond to functions \re{1.10} - blue solid line and \re{3.32} - with $n_{\rm max}=10$ red dashed line.}
\label{frequency}
\ec
\end{figure}

Now we evaluate the effect of ${\bm B}_{1}^{\dagger}(t){\bm B}_{2}(t)$ on the time-dependent
flux as a function of the Zeeman coupling.
For this purpose we use the Fourier series in the time domain and define:
\bea
J(t)\equiv \sqrt{\frac{2}{T_{2}-T_{1}}}\sum_{n=1}^{n_{\max}}J_{n}\sin\frac{\pi n(t-T_{1})}{T_{2}-T_{1}},
\lab{3.32}
\eea
with the coefficients $J_{n}$
\bea
J_{n}=\sqrt{\frac{2}{T_{2}-T_{1}}}\int_{T_{1}}^{T_{2}}J(x_{0},t)\sin\frac{\pi n(t-T_{1})}{T_{2}-T_{1}}dt,
\eea
where $[T_{1},T_{2}],\ (T_{1}=0,\ T_{2}=30)$ is the full collision time interval.
For summation limit $n_{\max}\rightarrow\infty$ the function \re{3.32} is $J(x_{0},t)$ in \re{1.10}.
To quantitatively describe the interference, first we filter out high-frequency Fourier components
from the time dependence by taking a smaller $n_{\max}$ (in our case $n_{\max}=10$)
limit in Eq.\re{3.32}. Now, the high-frequency terms do not contribute,
and in Fig. \ref{frequency} we see that plot of the function \re{1.10} symmetrically
oscillates around the plot of the filtered function \re{3.32}. The maximal
amplitude of oscillation is obtained for $\alpha=0$ and $\Delta=0$. As a result,
one can define the value corresponding to the strongest interference as:
\be
F_{\max}=\int_{T_{1}}^{T_{2}}\left[J(x_{0},t)-J(t)\right]^2dt, \qquad (\Delta=0,\alpha=0)
\lab{3.33}
\ee
having the value of $\approx 0.04$ at given system parameters.
The efficiency of the interference as a function of $\Delta$
is characterized by:
\bea
F(\Delta,\alpha)=\frac{1}{F_{\max}}\int_{T_{1}}^{T_{2}}\left[J(x_{0},t)-J(t)\right]^2dt,
\lab{3.34}
\eea
and the contribution of the backflow is evaluated as:
\bea
\Gamma(\Delta,\alpha)=\frac{2}{F_{\max}}\int_{T_{1}}^{T_{2}}J(t)\left[|J(x_{0},t)|-J(x_{0},t)\right]dt.
\lab{3.35}
\eea

The evaluation of interference and backflow, \re{3.34} and \re{3.35}
dependent on $\Delta$ is plotted in Fig. \ref{interference}
for given values of spin-orbit coupling  \re{2.21a} and \re{2.21}.
Fig. \ref{interference} and \ref{frequency} show that it is possible
to control the interference of two condensates using the spin-orbit coupling and synthetic magnetic field.
For strong field spins of particles are frozen in one direction and interference
is maximal. The zero value of the function $\Gamma(\Delta,\alpha)$ corresponds to the absence of backflow,
where the flux $J(x_{0},t)>0$ for any $t$. As one can see, the intervals of its zero
values are relatively small, meaning that the backflow is robust
against the spin-dependent interactions. {Figure \ref{interference} shows
that for the given system parameters the backflow disappears
if the interference parameter $F(\Delta,\alpha)$ is less than 0.5.}

As for the role of the interactions, they do not influence the momentum
of the packet, so they do not change its mean spin precession rate, affecting
the spins only marginally. However they
do influence the packet width and can prevent collision if they are strong
enough. To avoid these effects in the regime $p_{1}w_{1}\gg\hbar$ and  $p_{2}w_{2}\gg\hbar$ it is sufficient
to satisfy the condition of small contribution of the interatomic repulsion into the packet width.
Since in the absence of repulsion the packet spreads with the rate of the order of
$\hbar/Mw_{j}$, the interaction energy per atom
should be less than $\hbar^{2}/Mw_{j}^{2}$ to satisfy this condition.
A good candidate for a very weakly interacting BEC is $^{7}\mbox{Li}$ ensemble,
although, to the best of our knowledge, spin-orbit coupling effects have not been reported for this isotope.

\section{III. Conclusions}
We have shown that the superposition of two freely moving spin-orbit coupled condensates
gives rise to interference effects strongly dependent on the spin state of the
condensates at the collision time. The interference - characterizing both
the density and the flux - is strong when the spins of the two condensates
are parallel, and it disappears when the spin states are orthogonal.
These effects can be clearly seen in time-of-flight experiments,
and are at reach with the current technology for ultracold atoms.
In addition, the system exhibits a spin-dependent quantum backflow
behavior, which is relatively robust against synthetic spin-orbit coupling and magnetic
field. The ability to control the interference by synthetic
spin-orbit coupling and magnetic field can be useful for investigating the
quantum properties of atomic condensates and for  {interference of macroscopic spin-orbit coupled BEC-based qubits for 
quantum information applications.}

\acknowledgments
This work was supported by the University of Basque Country UPV/EHU under
program UFI 11/55, Spanish MEC (FIS2012-36673-C03-01 and  FIS2012-36673-C03-03),
and Grupos Consolidados UPV/EHU del Gobierno Vasco (IT-472-10). S.M. acknowledges EU-funded Erasmus Mundus Action 2 eASTANA,
"evroAsian Starter for the Technical Academic Programme" (Agreement No. 2001-2571/001-001-EMA2). M.P.
is supported by the Doctoral Scholarship of the University of Basque Country UPV/EHU.

\end{document}